\pdfoutput=1
\documentclass[runningheads]{llncs}

\title{Short Paper: Accountable Safety Implies Finality}
\author{%
Joachim Neu%
\and%
Ertem Nusret Tas%
\and%
David Tse%
}%
\institute{%
Stanford University\\%
\email{\{jneu,nusret,dntse\}@stanford.edu}%
}%

\usepackage{import}
\usepackage[utf8]{inputenc}

\usepackage{ifthen}
\usepackage{xifthen}   %
\usepackage{ifthenx}
\newcommand{\ifequal}[3]{\ifthenelse{\equal{#1}{#2}}{#3}{}}

\PassOptionsToPackage{hyphens}{url}%
\usepackage{hyperref}
\usepackage{graphicx}

\usepackage{amsmath}
\usepackage{amssymb}
\usepackage{amsfonts}
\usepackage{mathtools}

\usepackage{shortsym}

\usepackage{IEEEtrantools}
\allowdisplaybreaks

\usepackage[shortlabels,inline]{enumitem}
\setlist[itemize]{leftmargin=1.25em}
\setlist[enumerate]{leftmargin=1.75em}

\usepackage{import}
\makeatletter
\renewcommand*\env@matrix[1][*\c@MaxMatrixCols c]{%
    \hskip -\arraycolsep
    \let\@ifnextchar\new@ifnextchar
    \array{#1}}
\makeatother

\DeclarePairedDelimiter\abs{\lvert}{\rvert}
\DeclarePairedDelimiter\len{\lvert}{\rvert}
\DeclarePairedDelimiter\norm{\lVert}{\rVert}

\makeatletter
\let\oldabs\abs
\def\abs{\@ifstar{\oldabs}{\oldabs*}}
\let\oldlen\len
\def\len{\@ifstar{\oldlen}{\oldlen*}}
\let\oldnorm\norm
\def\norm{\@ifstar{\oldnorm}{\oldnorm*}}
\makeatother

\newcommand{\poly}[0]{\ensuremath{\operatorname{poly}}}

\usepackage{pifont}

\usepackage{dsfont}

\usepackage{xspace}

\newcommand{\cf}[0]{cf.\xspace}
\newcommand{\ie}[0]{\emph{i.e.}\xspace}
\newcommand{\eg}[0]{\emph{e.g.}\xspace}

\usepackage{xcolor}

\definecolor{mySUCardinalRed}{HTML}{8c1515}
\definecolor{mySUCardinalRedLight}{HTML}{B83A4B}
\definecolor{mySUCardinalRedDark}{HTML}{820000}
\definecolor{mySUWhite}{HTML}{ffffff}
\definecolor{mySUCoolGrey}{HTML}{53565A}
\definecolor{mySUBlack}{HTML}{2e2d29}
\definecolor{mySUBlack100}{HTML}{2e2d29}
\definecolor{mySUBlack90}{HTML}{43423E}
\definecolor{mySUBlack80}{HTML}{585754}
\definecolor{mySUBlack70}{HTML}{6D6C69}
\definecolor{mySUBlack60}{HTML}{767674}
\definecolor{mySUBlack50}{HTML}{979694}
\definecolor{mySUBlack40}{HTML}{ABABA9}
\definecolor{mySUBlack30}{HTML}{C0C0BF}
\definecolor{mySUBlack20}{HTML}{D5D5D4}
\definecolor{mySUBlack10}{HTML}{EAEAEA}

\definecolor{mySUPaloAlto}{HTML}{175E54}
\definecolor{mySUPaloAltoLight}{HTML}{2D716F}
\definecolor{mySUPaloAltoDark}{HTML}{014240}
\definecolor{mySUPaloVerde}{HTML}{279989}
\definecolor{mySUPaloVerdeLight}{HTML}{59B3A9}
\definecolor{mySUPaloVerdeDark}{HTML}{017E7C}
\definecolor{mySUOlive}{HTML}{8F993E}
\definecolor{mySUOliveLight}{HTML}{A6B168}
\definecolor{mySUOliveDark}{HTML}{7A863B}
\definecolor{mySUBay}{HTML}{6FA287}
\definecolor{mySUBayLight}{HTML}{8AB8A7}
\definecolor{mySUBayDark}{HTML}{417865}
\definecolor{mySUSky}{HTML}{4298B5}
\definecolor{mySUSkyLight}{HTML}{67AFD2}
\definecolor{mySUSkyDark}{HTML}{016895}
\definecolor{mySULagunita}{HTML}{007C92}
\definecolor{mySULagunitaLight}{HTML}{009AB4}
\definecolor{mySULagunitaDark}{HTML}{006B81}
\definecolor{mySUPoppy}{HTML}{E98300}
\definecolor{mySUPoppyLight}{HTML}{F9A44A}
\definecolor{mySUPoppyDark}{HTML}{D1660F}
\definecolor{mySUSpirited}{HTML}{E04F39}
\definecolor{mySUSpiritedLight}{HTML}{F4795B}
\definecolor{mySUSpiritedDark}{HTML}{C74632}
\definecolor{mySUIlluminating}{HTML}{FEDD5C}
\definecolor{mySUIlluminatingLight}{HTML}{FFE781}
\definecolor{mySUIlluminatingDark}{HTML}{FEC51D}
\definecolor{mySUPlum}{HTML}{620059}
\definecolor{mySUPlumLight}{HTML}{734675}
\definecolor{mySUPlumDark}{HTML}{350D36}
\definecolor{mySUBrick}{HTML}{651C32}
\definecolor{mySUBrickLight}{HTML}{7F2D48}
\definecolor{mySUBrickDark}{HTML}{42081B}
\definecolor{mySUArchway}{HTML}{5D4B3C}
\definecolor{mySUArchwayLight}{HTML}{766253}
\definecolor{mySUArchwayDark}{HTML}{2F2424}
\definecolor{mySUStone}{HTML}{7F7776}
\definecolor{mySUStoneLight}{HTML}{D4D1D1}
\definecolor{mySUStoneDark}{HTML}{544948}
\definecolor{mySUFog}{HTML}{DAD7CB}
\definecolor{mySUFogLight}{HTML}{F4F4F4}
\definecolor{mySUFogDark}{HTML}{B6B1A9}

\definecolor{mySUDigitalRed}{HTML}{B1040E}
\definecolor{mySUDigitalRedLight}{HTML}{E50808}
\definecolor{mySUDigitalRedDark}{HTML}{820000}
\definecolor{mySUDigitalBlue}{HTML}{006CB8}
\definecolor{mySUDigitalBlueLight}{HTML}{6FC3FF}
\definecolor{mySUDigitalBlueDark}{HTML}{00548f}
\definecolor{mySUDigitalGreen}{HTML}{008566}
\definecolor{mySUDigitalGreenLight}{HTML}{1AECBA}
\definecolor{mySUDigitalGreenDark}{HTML}{006F54}

\definecolor{myParula01Blue}{RGB}{0,114,189}
\definecolor{myParula02Orange}{RGB}{217,83,25}
\definecolor{myParula03Yellow}{RGB}{237,177,32}
\definecolor{myParula04Purple}{RGB}{126,47,142}
\definecolor{myParula05Green}{RGB}{119,172,48}
\definecolor{myParula06LightBlue}{RGB}{77,190,238}
\definecolor{myParula07Red}{RGB}{162,20,47}

\usepackage{tikz}
\usetikzlibrary{calc}
\usetikzlibrary{arrows}
\usetikzlibrary{arrows.meta}
\usetikzlibrary{patterns}
\usetikzlibrary{positioning}
\usetikzlibrary{decorations.pathreplacing}
\usetikzlibrary{calligraphy}
\usetikzlibrary{shapes.misc}
\usetikzlibrary{shapes.symbols}
\usetikzlibrary{shapes.arrows}
\usetikzlibrary{spy}

\usepackage{pgfplots}
\pgfplotsset{compat=1.18}
\usepgfplotslibrary{fillbetween}

\pgfplotsset{
    discard if not/.style 2 args={
            x filter/.code={
                    \edef\tempa{\thisrow{#1}}
                    \edef\tempb{#2}
                    \ifx\tempa\tempb
                    \else
                        
                    \fi
                }
        },
}

\tikzset{myparula11/.style={color=myParula01Blue,solid,mark=+,mark options={solid}}}
\tikzset{myparula12/.style={color=myParula01Blue,densely dashed,mark=x,mark options={solid}}}
\tikzset{myparula13/.style={color=myParula01Blue,densely dotted,mark=o,mark options={solid}}}
\tikzset{myparula14/.style={color=myParula01Blue,dashdotted,mark=triangle,mark options={solid}}}
\tikzset{myparula15/.style={color=myParula01Blue,dashdotdotted,mark=square,mark options={solid}}}

\tikzset{myparula21/.style={color=myParula02Orange,solid,mark=+,mark options={solid}}}
\tikzset{myparula22/.style={color=myParula02Orange,densely dashed,mark=x,mark options={solid}}}
\tikzset{myparula23/.style={color=myParula02Orange,densely dotted,mark=o,mark options={solid}}}
\tikzset{myparula24/.style={color=myParula02Orange,dashdotted,mark=triangle,mark options={solid}}}
\tikzset{myparula25/.style={color=myParula02Orange,dashdotdotted,mark=square,mark options={solid}}}

\tikzset{myparula31/.style={color=myParula03Yellow,solid,mark=+,mark options={solid}}}
\tikzset{myparula32/.style={color=myParula03Yellow,densely dashed,mark=x,mark options={solid}}}
\tikzset{myparula33/.style={color=myParula03Yellow,densely dotted,mark=o,mark options={solid}}}
\tikzset{myparula34/.style={color=myParula03Yellow,dashdotted,mark=triangle,mark options={solid}}}
\tikzset{myparula35/.style={color=myParula03Yellow,dashdotdotted,mark=square,mark options={solid}}}

\tikzset{myparula41/.style={color=myParula04Purple,solid,mark=+,mark options={solid}}}
\tikzset{myparula42/.style={color=myParula04Purple,densely dashed,mark=x,mark options={solid}}}
\tikzset{myparula43/.style={color=myParula04Purple,densely dotted,mark=o,mark options={solid}}}
\tikzset{myparula44/.style={color=myParula04Purple,dashdotted,mark=triangle,mark options={solid}}}
\tikzset{myparula45/.style={color=myParula04Purple,dashdotdotted,mark=square,mark options={solid}}}

\tikzset{myparula51/.style={color=myParula05Green,solid,mark=+,mark options={solid}}}
\tikzset{myparula52/.style={color=myParula05Green,densely dashed,mark=x,mark options={solid}}}
\tikzset{myparula53/.style={color=myParula05Green,densely dotted,mark=o,mark options={solid}}}
\tikzset{myparula54/.style={color=myParula05Green,dashdotted,mark=triangle,mark options={solid}}}
\tikzset{myparula55/.style={color=myParula05Green,dashdotdotted,mark=square,mark options={solid}}}

\tikzset{myparula61/.style={color=myParula06LightBlue,solid,mark=+,mark options={solid}}}
\tikzset{myparula62/.style={color=myParula06LightBlue,densely dashed,mark=x,mark options={solid}}}
\tikzset{myparula63/.style={color=myParula06LightBlue,densely dotted,mark=o,mark options={solid}}}
\tikzset{myparula64/.style={color=myParula06LightBlue,dashdotted,mark=triangle,mark options={solid}}}
\tikzset{myparula65/.style={color=myParula06LightBlue,dashdotdotted,mark=square,mark options={solid}}}

\tikzset{myparula71/.style={color=myParula07Red,solid,mark=+,mark options={solid}}}
\tikzset{myparula72/.style={color=myParula07Red,densely dashed,mark=x,mark options={solid}}}
\tikzset{myparula73/.style={color=myParula07Red,densely dotted,mark=o,mark options={solid}}}
\tikzset{myparula74/.style={color=myParula07Red,dashdotted,mark=triangle,mark options={solid}}}
\tikzset{myparula75/.style={color=myParula07Red,dashdotdotted,mark=square,mark options={solid}}}

\pgfplotsset{
    mysimpleplot/.style = {
            every axis plot/.prefix style={thick},
            width=1.05\linewidth,
            height=0.75\linewidth,
            title style={font=\footnotesize,align=center},
            legend cell align=left,
            legend style={font=\footnotesize},
            legend columns=3,
            legend style={
                    at={(0.5,1)},
                    yshift=0.3em,
                    anchor=south,
                    draw=none,
                    /tikz/every even column/.append style={
                            column sep=0.3em
                        },
                    cells={
                            align=left
                        }
                },
            grid=both,
            minor tick num=4,
            major grid style={solid,draw=gray!50},
            minor grid style={densely dotted,draw=gray!50},
            label style={font=\footnotesize,align=center},
            tick label style={font=\footnotesize},
        },
}

\pgfplotsset{
    myresultsplot01/.style={
            legend style={
                    at={(0.5,1.15)},
                    anchor=north,
                    legend columns=-1,
                    draw=none,
                    /tikz/every even column/.append style={column sep=1em},
                    cells={align=left},
                    font=\small,
                },
            enlarge x limits=0.15,
            ybar,
            bar width=6mm,
            xtick=data,
            width=\linewidth,
            height=0.6\linewidth,
            ymin=0,
            grid=both,
            minor y tick num=4,
            minor x tick num=1,
            major grid style={solid,draw=gray!50},
            minor grid style={densely dotted,draw=gray!50},
            label style={font=\small},
            tick label style={font=\small},
        },
}

\pgfplotsset{
    mysimpleresilienceplot01/.style = {
            mysimpleplot,
            ylabel={Adversary resilience $\beta$},
            height=0.45\linewidth,
            width=\linewidth,
            ymin=0,ymax=0.5,
            ytick={0,0.1,0.2,0.3,0.4,0.5},
            xmin=1e-5,xmax=1e2,
            grid=major,
        },
}

\pgfplotsset{
    mybandwidthplot01/.style = {
            mysimpleplot,
            ylabel={Minimum bandwidth $C$ (Mbps)},
            xlabel={Adversary resilience $\beta$},
            height=0.6\linewidth,
            width=\linewidth,
            xmin=0,xmax=0.5,
            xtick={0,0.1,0.2,0.3,0.4,0.5},
        },
}

\tikzset{blockchainold/.style={
            x=1.5cm,
            y=0.6cm,
            node distance=0.5cm,
            block/.style = {
                    minimum width=0.25cm,
                    minimum height=0.25cm,
                    draw,
                    shade,
                    top color=white,
                    bottom color=black!10,
                },
            block-adv1/.style = {
                    block,
                    bottom color=myParula01Blue!50,
                    draw=myParula01Blue!50!black
                },
            block-adv2/.style = {
                    block,
                    bottom color=myParula07Red!50,
                    draw=myParula07Red!50!black,
                },
            block-adv3/.style = {
                    block,
                    bottom color=myParula05Green!50,
                    draw=myParula05Green!50!black,
                },
            block-green/.style = {
                    block,
                    bottom color=myParula05Green!50,
                    draw=myParula05Green!50!black,
                },
            block-red/.style = {
                    block,
                    bottom color=myParula07Red!50,
                    draw=myParula07Red!70!black,
                },
            block-gray/.style = {
                    block,
                    bottom color=black!30,
                },
            block-big/.style = {
                    minimum width=0.7cm,
                    minimum height=0.7cm,
                    draw,
                    shade,
                    top color=white,
                    bottom color=black!10,
                },
            branch/.style = {
                    minimum width=0.1cm,
                    minimum height=0.1cm,
                    draw,
                    circle,
                    inner sep=0,
                    fill=black,
                },
            link/.style = {
                    -latex,
                },
            hiddenlink/.style = {
                    dashed,
                },
            hiddenlink-adv1/.style = {
                    hiddenlink,
                    draw=myParula01Blue!50!black,
                },
            hiddenlink-adv2/.style = {
                    hiddenlink,
                    draw=myParula07Red!50!black,
                },
            hiddenlink-adv3/.style = {
                    hiddenlink,
                    draw=myParula05Green!50!black,
                },
            label/.style = {
                },
            label-adv1/.style = {
                    label,
                    text=myParula01Blue!50!black,
                },
            label-adv2/.style = {
                    label,
                    text=myParula07Red!50!black,
                },
            label-adv3/.style = {
                    label,
                    text=myParula05Green!50!black,
                },
        }
}

\tikzset{blockchain/.style={
            x=0.5cm,
            y=0.55cm,
            node distance=0.5cm,
            block/.style = {
                    minimum width=0.3cm,
                    minimum height=0.3cm,
                    draw,
                    shade,
                    top color=white,
                    bottom color=black!10,
                    inner sep=0,
                },
            block-adv/.style = {
                    block,
                    bottom color=myParula07Red!50,
                    draw=myParula07Red!50!black,
                },
            block-hon/.style = {
                    block,
                    bottom color=myParula05Green!50,
                    draw=myParula05Green!50!black,
                },
            block-blank/.style = {
                    minimum width=0.3cm,
                    minimum height=0.3cm,
                    rounded corners,
                    inner sep=0,
                },
            link/.style = {
                    -latex,
                },
            link-adv/.style = {
                    link,
                },
            link-hon/.style = {
                    link,
                },
        }
}

\usepackage{multirow,booktabs}
\usepackage{makecell}

\usepackage{algorithm}
\usepackage{algorithmicx}
\usepackage[noend]{algpseudocode}

\algnewcommand{\algorithmicswitch}{\textbf{switch}}
\algdef{SE}[SWITCH]{Switch}{EndSwitch}[1]{\algorithmicswitch\ #1\ \algorithmicdo}{\algorithmicend\ \algorithmicswitch}%
\algtext*{EndSwitch}%

\algnewcommand{\algorithmiccase}{\textbf{case}}
\algdef{SE}[CASE]{Case}{EndCase}[1]{\algorithmiccase\ #1}{\algorithmicend\ \algorithmiccase}%
\algtext*{EndCase}%

\algnewcommand{\algorithmicon}{\textbf{on}}
\algdef{SE}[ON]{On}{EndOn}[1]{\algorithmicon\ #1\ \algorithmicdo}{\algorithmicend\ \algorithmicon}%
\algtext*{EndOn}%

\algnewcommand{\algorithmicat}{\textbf{at}}
\algdef{SE}[AT]{At}{EndAt}[1]{\algorithmicat\ #1\ \algorithmicdo}{\algorithmicend\ \algorithmicat}%
\algtext*{EndAt}%

\algnewcommand{\algorithmicrealfunction}{\textbf{function}}
\algdef{SE}[REALFUNCTION]{RealFunction}{EndRealFunction}[1]{\algorithmicrealfunction\ #1\ \algorithmicdo}{\algorithmicend\ \algorithmicrealfunction}%
\algtext*{EndRealFunction}%

\algnewcommand{\algorithmicthroughout}{\textbf{throughout}}
\algdef{SE}[Throughout]{Throughout}{EndThroughout}[1]{\algorithmicthroughout\ #1\ \algorithmicdo}{\algorithmicend\ \algorithmicthroughout}%
\algtext*{EndThroughout}%

\algnewcommand{\algorithmicforever}{\textbf{forever}}
\algdef{SE}[FOREVER]{Forever}{EndForever}[0]{\algorithmicforever\ \algorithmicdo}{\algorithmicend\ \algorithmicforever}%
\algtext*{EndForever}%

\algrenewcommand{\algorithmicdo}{}
\algrenewcommand{\algorithmicthen}{}

\algnewcommand{\algorithmicgoto}{\textbf{goto}}%
\algnewcommand{\Goto}[1]{\algorithmicgoto~\ref{#1}}%

\algnewcommand{\algorithmicassert}{\textbf{assert}}%
\algnewcommand{\Assert}[1]{\algorithmicassert~{#1}}%

\algnewcommand{\algorithmicbreak}{\textbf{break}}%
\algnewcommand{\Break}[0]{\algorithmicbreak}%

\algnewcommand{\algorithmicwaiton}{\textbf{wait on}}%
\algnewcommand{\WaitOn}[1]{\algorithmicwaiton~{#1}}%

\algnewcommand{\LineComment}[1]{\State \(\triangleright\) \textit{#1}}
\algnewcommand{\InlineRequire}[1]{\textbf{require} {#1}}

\newcommand{\GST}[0]{\ensuremath{\mathsf{GST}}}

\newcommand{\cl}[0]{\ensuremath{\mathsf{cl}}}

\newcommand{\tx}[0]{\ensuremath{\mathsf{tx}}}

\newcommand{\Env}[0]{\CZ}
\newcommand{\Adv}[0]{\CA}

\newcommand{\LOG}[2]{\ensuremath{\mathsf{LOG}_{#1}^{#2}}}

\newcommand{\asr}{accountable-safety resilience\xspace}

\newcommand{\ASR}{Accountable-Safety Resilience\xspace}

\newcommand{\GAT}[0]{\ensuremath{\mathsf{GAT}}}

\newcommand{\replicam}[0]{\ensuremath{P}}

\newcommand{\Tconfirm}[0]{\ensuremath{T_{\mathrm{conf}}}}

\newcommand{\betaS}[0]{\ensuremath{\beta_{\mathrm{s}}}}
\newcommand{\betaA}[0]{\ensuremath{\beta_{\mathrm{a}}}}
\newcommand{\betaF}[0]{\ensuremath{\beta_{\mathrm{f}}}}
\newcommand{\betaL}[0]{\ensuremath{\beta_{\mathrm{l}}}}

\newcommand{\fS}[0]{\ensuremath{f_{\mathrm{s}}}}
\newcommand{\fA}[0]{\ensuremath{f_{\mathrm{a}}}}
\newcommand{\fL}[0]{\ensuremath{f_{\mathrm{l}}}}

\newcommand{\AdvEnvSync}[0]{\ensuremath{(\Adv_{\mathrm{s}}, \Env_{\mathrm{s}})}}
\newcommand{\AdvEnvPsync}[0]{\ensuremath{(\Adv_{\mathrm{p}}, \Env_{\mathrm{p}})}}
\newcommand{\AdvEnvInstant}[0]{\ensuremath{(\Adv_{\mathrm{i}}, \Env_{\mathrm{i}})}}
\newcommand{\AdvEnvPG}[0]{\ensuremath{(\Adv_{\mathrm{pda}}, \Env_{\mathrm{pda}})}}
\newcommand{\AdvInstantI}[0]{\ensuremath{\Adv^{(i)}_{\mathrm{s}}}}
\newcommand{\AdvPsync}[0]{\ensuremath{\Adv_{\mathrm{p}}}}
\newcommand{\AdvPG}[0]{\ensuremath{\Adv_{\mathrm{pda}}}}

\usepackage{tango-icons}
\usepackage{noto-emoji-easy}

\newcommand{\blfootnote}[1]{%
\begingroup%
\renewcommand\thefootnote{}\footnotetext{#1}%
\endgroup%
}

\newcommand{\myparagraph}[1]{\smallskip\noindent\emph{#1}\hspace{0.5em}}
\newcommand{\myparagraphbf}[1]{\smallskip\noindent\textbf{#1}\hspace{0.5em}}

\begin{document}
\maketitle%
\blfootnote{JN and ENT are listed alphabetically.}%
\begin{abstract}%
    Motivated by proof-of-stake (PoS) blockchains such as Ethereum,
    two key desiderata have recently been studied
    for Byzantine-fault tolerant (BFT)
    state-machine replication (SMR) consensus protocols:
    \emph{Finality} means that the protocol
    retains consistency,
    as long as less than a certain fraction of validators are malicious,
    even in \emph{partially-synchronous} environments
    that allow for 
    temporary violations of
    assumed network delay bounds.
    \emph{Accountable safety} means that
    in any case of inconsistency,
    a certain fraction of validators can be \emph{identified} to have
    \emph{provably} violated the protocol.
    Earlier works have developed impossibility results
    and protocol constructions for
    these properties separately.
    We show that accountable safety implies finality,
    thereby unifying
    earlier results.
\end{abstract}

\section{Introduction}
\label{sec:introduction}

\myparagraph{Consensus.}
The purpose of a consensus protocol for state-machine replication (SMR) is for a set of parties to reach agreement on how to sequence incoming transactions into a linear order called a \emph{ledger}. This task is non-trivial because communication between parties might be delayed, and some parties might deviate from the protocol in an arbitrary manner (\emph{Byzantine faults}) with the goal to undermine consensus. A consensus protocol is \emph{secure} if even in the presence of these disturbances, it guarantees two complementary properties:
\emph{safety}, meaning that the ledgers output by non-faulty parties across time are consistent,
and \emph{liveness}, meaning that transactions make it to the output ledger `soon'.
The protocol is then called Byzantine-fault tolerant (BFT), and the fraction of faulty parties it can tolerate while remaining secure is called its \emph{resilience}.

\myparagraph{Finality.}
This basic formulation has 
been extended in two directions.
On the one hand, while some early consensus protocols~\cite{dolevstrong} assume that network communication \emph{always} obeys a \emph{known} delay upper-bound (\ie, \emph{synchronous network}~\cite{sync-model}), later constructions~\cite{psync-model,pbft} pushed to strengthen security to ensure consistency also under temporary network delay-bound violations (\ie, \emph{partially-synchronous network}~\cite{psync-model}). Such periods of asynchrony might be caused, for instance, by temporary network partitions. The strengthened safety property that ensures consistency also under periods of asynchrony is called \emph{finality}~\cite{casper,ebbandflow}.

\myparagraph{Accountable Safety.}
On the other hand, consensus protocols for proof-of-stake (PoS) blockchains such as Ethereum seek to strengthen safety to enable \emph{accountability}~\cite{casper,gasper,aadilemma,ebbandflow,bftforensics,snapandchat,polygraph,revisitingtendermint,peerreview,faultdetectionproblem,blockchainisdead}. In permissionless blockchains, parties are no longer inanimate computers which might exhibit technical faults but are otherwise aligned under the control of one organizational entity. 
Instead, parties are controlled by different mutually distrusting self-interested players that might deviate from the protocol if they expect a profit from doing so. In this setting, it was proposed to strengthen safety to \emph{accountable safety}, where besides ensuring consistency up to some adversarial resilience, if the adversary exceeds that resilience
and causes a safety violation,
then a certain fraction of parties can also be identified to have provably violated the protocol.
\myparagraph{Prior Works.}
Various works have studied the fundamental limits of finality~\cite{flpimpossibility,psync-model,captheorem} and of accountable safety~\cite{bftforensics}, as well as their relationships to other desiderata such as liveness under dynamic participation~\cite{blockchain-cap-theorem,aadilemma}, and how protocols can be constructed that achieve various combinations of these properties~\cite{ebbandflow,aadilemma}.
While finality and accountable safety `feel similar', characterizing their exact relation has remained open. For instance, some protocols provide finality but do not provide accountable safety~\cite{bftforensics,aadilemmafull}.
On the other hand, 
we readily observe
(details in Section~\ref{sec:finality-psync}) that
an additional round of voting to `checkpoint'
the output ledger of a consensus protocol\footnote{%
A `lock and vote' add-on was contemporaneously used
for the unrelated problems
of resilience-optimal flexible consensus~\cite{oflex}
and of using Bitcoin as a staking asset~\cite{babylonlightpaper}.}
designed for synchronous networks can also be used to upgrade that ledger to provide accountable safety,
but does not yield a protocol for partially-synchronous networks.
In particular, the so-extended protocol may not recover from liveness faults 
induced during a period of asynchrony (Section~\ref{sec:finality-psync}).

\myparagraph{Main Result.}
We show that 
accountable safety implies finality (Section~\ref{sec:acc-implies-fin}).
To this end, on a high level,
we show that for any given protocol,
if there exists
an adversary strategy that leads to a safety violation
under partial synchrony,
then there exists an adversary strategy
that leads to a safety violation
but not enough adversary parties can be
identified as protocol violators,
even if the network is delay-free, \ie, messages arrive instantly.
Intuitively, the
more constraints there are on network delays,
the easier it is for a protocol to guarantee accountable safety.
Our argument shows that even the weakest form of accountable safety,
namely for delay-free networks,
is still so strong that it implies finality.

Our result
unifies prior works and directs future work (Section~\ref{sec:alternative-proofs}):
The availability--accountability dilemma~\cite{aadilemma}
turns out to be implied by
the availability--finality dilemma~\cite{ebbandflow},
a blockchain-variant of 
the CAP theorem~\cite{captheorem,blockchain-cap-theorem}.
Impossibilities for
accountability and liveness resiliences~\cite{bftforensics}
turn out to be implied by 
impossibilities
for consensus under partial synchrony~\cite{psync-model}.
Upgrading a ledger to provide accountable safety~\cite{aadilemma,snapandchat}
implies adding finality~\cite{casper,checkpointedlc,grandpa,afgjort,nearpost,hfg}.

\section{Model}
\label{sec:model}

\myparagraph{Notation.}
For $m\in\IN$, let $[m] \triangleq \{1,2,...,m\}$.
An event happens with probability \emph{negligible} in the security parameter $\lambda$
if its probability is $o(1/\poly(\lambda))$.

\myparagraph{Replicas and Clients.}
SMR consensus protocols have two types of participants:
\emph{replicas} and \emph{clients}.
Replicas are input \emph{transactions} by the environment, 
interact with each other towards agreeing on how these transactions should be ordered,
and make some protocol messages (\eg, blocks, votes) available to clients upon request.
\emph{Clients} query replicas for these protocol messages, 
and, upon collecting messages from a sufficiently large subset of replicas, 
output a sequence of transactions called the \emph{ledger} and denoted by $\LOG{}{}$.
The goal of the SMR protocol is to ensure that clients agree on a single ever-growing
transaction sequence.

\myparagraph{Environment and Adversary.}
Each of the $n$ replicas has a (unique) cryptographic identity that is 
known
to all parties.
Up to $f$ replicas can be corrupted at the beginning of the protocol execution
by a computationally-bounded \emph{adversary} $\Adv$,
which then obtains the internal state of these replicas,
and can make them deviate from the protocol in arbitrary ways (Byzantine faults).
The remaining $(n-f)$ replicas are \emph{honest} and follow the protocol as specified.

Time proceeds in slots.
Replicas can exchange messages, subject to 
adversary delays.
We consider a \emph{partially-synchronous network} with adversary-environment tuple $\AdvEnvPsync$,
where
the adversary can delay messages 
arbitrarily until a
\emph{global stabilization time} $\GST$ that can be chosen adaptively by the adversary.
After $\GST$, $\AdvPsync$ has to deliver messages within a delay upper-bound of $\Delta$
which is known to the protocol.
If $\GST$ is known and zero, then the network is said to be \emph{synchronous}, and denoted by $\AdvEnvSync$.
Furthermore,
a network is called \emph{delay-free}
and denoted by $\AdvEnvInstant$
if all messages 
reach their recipients instantaneously, \ie, the network is synchronous with $\Delta = 0$.
The three network models are ordered in the sense that
from partial synchrony to synchrony to delay-freeness,
for fixed $\Delta$,
the adversary's capabilities are strictly increasingly constrained.

\myparagraph{Safety and Liveness Resiliences.}
Let $\LOG{t}{\cl}$ denote the ledger in the view of a client $\cl$ at time slot $t$.
\begin{definition}
    \label{def:security}
    A consensus protocol 
    is \emph{secure} with confirmation time $\Tconfirm$ iff:
    \vspace{-0.5em}
    \begin{itemize}
        \item \emph{Safety:} For all $t,t'$ and $\cl,\cl'$, either $\LOG{t}{\cl}$ is a prefix of $\LOG{t'}{\cl'}$, or vice versa.
        \item \emph{Liveness:} If some $\tx$ is input to an honest replica by some $t$, then, for all $t' \geq \max(t,\GST)+\Tconfirm$, and all $\cl$, $\tx \in \LOG{t'}{\cl}$.
    \end{itemize}
\end{definition}
A protocol is said to provide \emph{$f$-safety ($f$-liveness)} if the protocol satisfies safety (liveness),
except with negligible probability, for any adversary controlling at most $f$ replicas.
Here, $f$ is the protocol's safety (liveness) \emph{resilience}.
\begin{definition}
    \label{def:finality}
    A consensus protocol satisfies \emph{$f$-finality} (\ie, is \emph{$f$-final}) 
    if it satisfies $f$-safety under a partially-synchronous network.
\end{definition}
Note that $f$-finality does not imply $f$-liveness after $\GST$.
An $f$-final protocol may not be secure under partial synchrony
due to liveness violations (\cf Section~\ref{sec:finality-psync}).

\myparagraph{\ASR.}
Building on 
\emph{$\alpha$-ac\-count\-a\-ble-safety}~\cite{casper,snapandchat}, the \asr of a protocol is defined
(see~\cite{aadilemma} for details on replica--client interaction and forensic algorithm, there called `adjudication function'):
\begin{definition}
    \label{def:accountable-safety}
    A consensus protocol provides \emph{accountable safety} 
    with resilience $\fA$ (\ie, is \emph{$\fA$-accountable safe}) 
    iff whenever there is a safety violation,
    except with negligible probability,
    (i) at least $\fA$ adversarial replicas are identified by a \emph{forensic algorithm} as protocol violators, and 
    (ii) no honest replica is identified.
\end{definition}
Specifically, if clients $\cl, \cl'$ at $t, t'$ disagree on the output ledger,
they exchange the protocol messages that led to their respective $\LOG{t}{\cl}, \LOG{t'}{\cl'}$.
Given a set of messages such that there are two subsets
based on which a client would confirm two conflicting ledgers,
each client can invoke the protocol's \emph{forensic algorithm}
to identify $\fA$ replicas 
who have provably violated the protocol~\cite{bftforensics}.
Note that 
the functioning of the forensic algorithm is not conditional on assumptions such as a fraction of replicas being honest, and it does not falsely accuse honest replicas.

\section{Accountability Implies Finality}
\label{sec:acc-implies-fin}

By Definition~\ref{def:accountable-safety}, if a protocol provides $(f+1)$-accountable safety under partial synchrony
(which is the strongest form of accountable safety, considering that 
among the models considered here,
the adversary's capabilities
are least constrained in the partially-synchronous model),
then it also satisfies safety under partial synchrony with up to $f$ adversary replicas, \ie, it is $f$-final.
(This is because if the number of adversary replicas is less than $f+1$, the forensic algorithm \emph{cannot} identify at least $f+1$ \emph{adversary} replicas, 
implying that the protocol must be safe.)
Perhaps more surprisingly, Theorem~\ref{thm:acc-implies-fin} below proves that if a protocol provides $(f+1)$-accountable safety in a delay-free network
(which is the weakest form of accountable safety, since the adversary's capabilities
are most constrained in the delay-free network model), 
it must still be the case that the protocol is $f$-final.
This immediately implies that for all
network delay models,
$(f+1)$-accountable safety 
of a protocol
implies $f$-finality for that protocol.%
\begin{theorem}
    \label{thm:acc-implies-fin}
    If a consensus protocol provides $(f+1)$-accountable safety in a delay-free network, then it also satisfies $f$-finality.
\end{theorem}

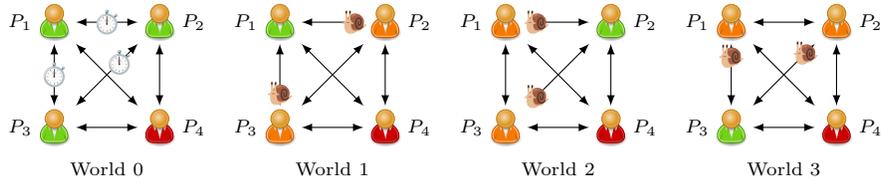
\begin{figure}[tb]
\centering
    \centering
    \begin{tikzpicture}[x=1cm,y=1cm]
        \scriptsize

        \begin{scope}[x=0.7cm,y=0.7cm]
        
            \node [inner sep=0] (P1) at (-1,+1) {\tangoicon[height=2em]{emblem_person_green}};
            \node [inner sep=0] (P2) at (+1,+1) {\tangoicon[height=2em]{emblem_person_green}};
            \node [inner sep=0] (P3) at (-1,-1) {\tangoicon[height=2em]{emblem_person_green}};
            \node [inner sep=0] (P4) at (+1,-1) {\tangoicon[height=2em]{emblem_person_red}};
            
            \node [anchor=east,inner sep=0] at (P1.west) {$P_1$};
            \node [anchor=west,inner sep=0] at (P2.east) {$P_2$};
            \node [anchor=east,inner sep=0] at (P3.west) {$P_3$};
            \node [anchor=west,inner sep=0] at (P4.east) {$P_4$};

            \draw [latex-latex] (P1) -- (P2) node [midway] {\notoemoji[height=1.25em]{stopwatch}};
            \draw [latex-latex] (P1) -- (P3) node [midway] {\notoemoji[height=1.25em]{stopwatch}};
            \draw [latex-latex] (P1) -- (P4);
            \draw [latex-latex] (P2) -- (P3) node [pos=0.3] {\notoemoji[height=1.25em]{stopwatch}};
            \draw [latex-latex] (P2) -- (P4);
            \draw [latex-latex] (P3) -- (P4);

            \node [yshift=-1.25cm] at (0,0) {World $0$};
            
        \end{scope}

        \begin{scope}[x=0.7cm,y=0.7cm,xshift=3cm]
        
            \node [inner sep=0] (P1) at (-1,+1) {\tangoicon[height=2em]{emblem_person_green}};
            \node [inner sep=0] (P2) at (+1,+1) {\tangoicon[height=2em]{emblem_person_orange}};
            \node [inner sep=0] (P3) at (-1,-1) {\tangoicon[height=2em]{emblem_person_orange}};
            \node [inner sep=0] (P4) at (+1,-1) {\tangoicon[height=2em]{emblem_person_red}};
            
            \node [anchor=east,inner sep=0] at (P1.west) {$P_1$};
            \node [anchor=west,inner sep=0] at (P2.east) {$P_2$};
            \node [anchor=east,inner sep=0] at (P3.west) {$P_3$};
            \node [anchor=west,inner sep=0] at (P4.east) {$P_4$};

            \draw [latex-latex] (P1) -- (P2) node [pos=0.85] {\notoemoji[height=1.25em]{snail}};
            \draw [latex-latex] (P1) -- (P3) node [pos=0.8] {\notoemoji[height=1.25em]{snail}};
            \draw [latex-latex] (P1) -- (P4);
            \draw [latex-latex] (P2) -- (P3);
            \draw [latex-latex] (P2) -- (P4);
            \draw [latex-latex] (P3) -- (P4);

            \node [yshift=-1.25cm] at (0,0) {World $1$};
            
        \end{scope}

        \begin{scope}[x=0.7cm,y=0.7cm,xshift=6cm]
        
            \node [inner sep=0] (P1) at (-1,+1) {\tangoicon[height=2em]{emblem_person_orange}};
            \node [inner sep=0] (P2) at (+1,+1) {\tangoicon[height=2em]{emblem_person_green}};
            \node [inner sep=0] (P3) at (-1,-1) {\tangoicon[height=2em]{emblem_person_orange}};
            \node [inner sep=0] (P4) at (+1,-1) {\tangoicon[height=2em]{emblem_person_red}};
            
            \node [anchor=east,inner sep=0] at (P1.west) {$P_1$};
            \node [anchor=west,inner sep=0] at (P2.east) {$P_2$};
            \node [anchor=east,inner sep=0] at (P3.west) {$P_3$};
            \node [anchor=west,inner sep=0] at (P4.east) {$P_4$};

            \draw [latex-latex] (P1) -- (P2) node [pos=0.15] {\notoemoji[height=1.25em]{snail}};
            \draw [latex-latex] (P1) -- (P3);
            \draw [latex-latex] (P1) -- (P4);
            \draw [latex-latex] (P2) -- (P3) node [pos=0.85] {\notoemoji[height=1.25em]{snail}};
            \draw [latex-latex] (P2) -- (P4);
            \draw [latex-latex] (P3) -- (P4);

            \node [yshift=-1.25cm] at (0,0) {World $2$};
            
        \end{scope}

        \begin{scope}[x=0.7cm,y=0.7cm,xshift=9cm]
        
            \node [inner sep=0] (P1) at (-1,+1) {\tangoicon[height=2em]{emblem_person_orange}};
            \node [inner sep=0] (P2) at (+1,+1) {\tangoicon[height=2em]{emblem_person_orange}};
            \node [inner sep=0] (P3) at (-1,-1) {\tangoicon[height=2em]{emblem_person_green}};
            \node [inner sep=0] (P4) at (+1,-1) {\tangoicon[height=2em]{emblem_person_red}};
            
            \node [anchor=east,inner sep=0] at (P1.west) {$P_1$};
            \node [anchor=west,inner sep=0] at (P2.east) {$P_2$};
            \node [anchor=east,inner sep=0] at (P3.west) {$P_3$};
            \node [anchor=west,inner sep=0] at (P4.east) {$P_4$};

            \draw [latex-latex] (P1) -- (P2);
            \draw [latex-latex] (P1) -- (P3) node [pos=0.2] {\notoemoji[height=1.25em]{snail}};
            \draw [latex-latex] (P1) -- (P4);
            \draw [latex-latex] (P2) -- (P3) node [pos=0.15] {\notoemoji[height=1.25em]{snail}};
            \draw [latex-latex] (P2) -- (P4);
            \draw [latex-latex] (P3) -- (P4);

            \node [yshift=-1.25cm] at (0,0) {World $3$};
            
        \end{scope}
        
    \end{tikzpicture}
    \vspace{-0.5em}
    \caption{Execution of a consensus protocol with four replicas. 
    World $0$ is partially-synchronous,
    worlds $1$, $2$ and $3$ are delay-free. 
    Red replica $\replicam_4$ is adversary in all
    worlds. 
    Orange replicas are adversary but do not violate the protocol rules other than delaying the sending/receiving of messages to/from the honest replica. 
    Green replicas are honest.}
\label{fig:example}
\end{figure}

\myparagraph{Intuition.}
For intuition,
consider the scenario with $n=4, f=1$.
We argue the equivalent claim that
without $1$-finality, there is no $2$-accountable safety.
For contradiction, consider a consensus protocol executed by four replicas $\replicam_i$, $i \in [4]$, that is not $1$-final, yet $2$-accountable safe under a delay-free network.

Consider the following executions:
In world $0$ (Figure~\ref{fig:example}), the network is partially-synchronous,
and $\replicam_4$ (red in Figure~\ref{fig:example}) is adversary.
Besides protocol deviations of $\replicam_4$,
the adversary delays messages among honest replicas to cause a safety violation
(which is possible because the protocol is assumed not $1$-final).
In worlds $1$, $2$, and $3$ (Figure~\ref{fig:example}), the network is delay-free,
and there are three adversary replicas.
The replica $\replicam_4$ is adversary, and in all of these worlds
behaves the same as in world $0$.
The remaining adversary replicas (orange in Figure~\ref{fig:example})
behave like their honest counterparts in world $0$, except they
\emph{emulate} the message delivery schedule an honest replica would have had
in their place in
world $0$,
by delaying the sending/receiving of messages to/from the honest replica.
Clients observing the protocol cannot distinguish between any of the worlds $0$, $1$, $2$, and $3$.
Therefore, there is a safety violation in worlds $1$, $2$, and $3$ as well.

Finally, since the protocol is assumed to be $2$-accountable safe in a delay-free network, 
the forensic algorithm called by the clients 
identifies $2$ replicas as protocol violators in each of the worlds $1$, $2$ and $3$.
However, as these worlds are indistinguishable, 
there is a non-negligible probability
that 
the forensic algorithm wrongly identifies an honest replica, which is a contradiction,
as desired.
\begin{proof}[of Theorem~\ref{thm:acc-implies-fin}]
    We prove the contrapositive.
    For contradiction, suppose a protocol does not satisfy $f$-finality, 
    yet provides $(f+1)$-accountable safety under a delay-free network.
    We consider a world $0$ with a partially-synchronous network, 
    and $n-f$ worlds indexed by $i \in [n-f]$ with delay-free networks.

    \noindent
    \textbf{World $0$:}
    Consider 
    clients $\cl$, $\cl'$ and $n$ replicas.
    The network is partially-synchronous and the adversary $\AdvPsync$ controls $f$ replicas, denoted by $\replicam_{n-f+1}, \ldots, \replicam_{n}$.
    The remaining replicas are honest.
    Safety is violated and the clients $\cl$ and $\cl'$ output conflicting ledgers with some non-negligible probability.

    \noindent
    \textbf{World $i$:}
    Consider 
    clients $\cl$, $\cl'$ and $n$ replicas.
    The network is delay-free and the adversary $\AdvInstantI$ of world $i$ corrupts all replicas except $\replicam_i$.
    The $f$ replicas that were adversary in world $0$ behave the same as in world $0$.
    The remaining adversary replicas behave like the corresponding honest replicas in world $0$,
    except they also emulate the network delay of world $0$:
    For each message sent by $\replicam_i$, adversary replicas pretend as if the message was delivered at the time slot
    in which it was delivered in world $0$,
    even though it was in fact delivered instantly in world $i$.
    Adversary replicas also send the same messages to $\replicam_i$ as in world $0$,
    but they ensure that these messages are delivered to $\replicam_i$ at the same time slots within world $i$ as
    they were delivered in world $0$, by delaying the sending 
    if necessary (after
    delayed sending, the delay-free network will deliver them instantly).

    As $\replicam_i$ receives the same messages at the same time slots
    in world $i$ and world $0$,
    it cannot distinguish the worlds,
    and
    $\replicam_i$ shows the same behavior in both.

    As the adversary replicas simulate their behavior from world $0$ within world $i$, 
    and $\replicam_i$ shows the same behavior in both worlds, 
    $\cl$ and $\cl'$ cannot distinguish the two worlds $0$ and $i$.
    Thus, they output conflicting ledgers with non-negligible probability.
    In this case, by the assumed $(f+1)$-accountable safety of the protocol under delay-free networks,
    the forensic algorithm, invoked with the information received by these clients from the replicas, identifies at least $f+1$ replicas as protocol violators in world $i$ with non-negligible probability.

    Finally, since the two worlds $0$ and $i$ are indistinguishable for $\cl$ and $\cl'$ for all $i$, 
    the worlds $i \in [n-f]$ are indistinguishable as well for $\cl$ and $\cl'$.
    Thus, as long as $n = O(\poly(\lambda))$, the forensic algorithm 
    has a non-negligible probability to falsely accuse an honest replica,
    which is a contradiction.
    \qed
\end{proof}

Theorem~\ref{thm:acc-implies-fin} holds even if 
the forensic algorithm were to know whether the network is partially-synchronous or delay-free (while we block clients and replicas from learning this).
In the proof above, indistinguishability between worlds $0$ and $i$ is used to infer that (i) safety is violated in world $i$ as it is in world $0$, and (ii) worlds $i,j>0$ are indistinguishable, and thus 
an honest replica is likely falsely accused.
The argument for (i) remains,
but since the forensic algorithm can now 
distinguish 
worlds $0$ and $i$,
we must argue for (ii) directly.
Indeed, as 
worlds $i,j>0$ are all delay-free, and 
this is the only extra information given to forensic algorithm,
worlds $i$, $j$ still cannot be distinguished by the forensic algorithm.

\section{Simplification of Earlier Results}
\label{sec:alternative-proofs}
\myparagraphbf{Impossibility of Finality $\implies$ Impossibility of Accountability.}
Earlier work~\cite[Theorem B.1]{bftforensics} shows that no protocol can be $\fA$-accountable safe and $\fL$-live for $2\fL + \fA > n$ under a synchronous or partially-synchronous network.
This result follows directly from Theorem~\ref{thm:acc-implies-fin}, combined with the safety--liveness bound under partial synchrony 
(\cite[Theorem 4.4]{psync-model})
restated below:
\begin{proposition}[From~\cite{psync-model}, see also~\cite{aadilemmafullv1,multithreshold,multithresholdbft}]
\label{prop:psync-bound}
No protocol can satisfy $\fS$-finality, and $\fL$-liveness under a delay-free network, for $2\fL + \fS \geq n$.
\end{proposition}
In other words, given $\fS$, $\fL$, $2\fL+\fS \geq n$, no protocol can simultaneously preserve its safety under asynchrony with $\fS$ adversary replicas, and remain live with $\fL$ adversary replicas, even if the network is delay-free.
This is stronger than the claim that no protocol provides $\fS$-safety and $\fL$-liveness under partial synchrony, yet this stronger result directly follows from the proof of \cite[Theorem 4.4]{psync-model}.
Combining Proposition~\ref{prop:psync-bound} with Theorem~\ref{thm:acc-implies-fin}, one readily obtains that no protocol can be $\fA$-accountable safe and $\fL$-live under a synchronous network for any $\Delta$, including a delay-free network, if $2\fL + \fA > n$:
\begin{corollary}
   \label{cor:acc-implies-fin-1}
   No protocol can be $\fA$-accountable safe and $\fL$-live, for $2\fL + \fA > n$, under a synchronous or partially-synchronous network, for any $\Delta$.
\end{corollary}

\myparagraphbf{Availability--Finality Dilemma $\implies$ Availability--Accountability Dilemma.}
To state the availability--accountability dilemma, we recall the formal model for dynamic participation 
(\ie, temporary crash faults)~\cite{aadilemma,sleepy}.
Before a \emph{global awake time}
$\GAT$, the adversary $\Adv$ can
determine for every honest replica and time slot,
whether the replica is \emph{awake} (\ie, online) or \emph{asleep} (\ie, offline) in that slot.
After $\GAT$, all honest replicas are awake.
Awake replicas follow the protocol.
Asleep replicas have a \emph{temporary crash fault},
and do not execute the protocol in the respective time slot.
Adversary replicas are always awake.
Messages sent to a replica while asleep are 
processed by
the replica
whenever it wakes up.
The adversary-environment tuple $\AdvEnvPG$ models a partially-synchronous network with $\GST < \infty$ and $\GAT \in [0, \infty)$ that can be adaptively chosen by $\AdvPG$ and which are not known by honest replicas or the protocol designer.
Let $\beta$ denote the largest \emph{fraction} across all time slots, of adversary replicas among awake replicas.
Below, $\betaS$-safety, $\betaL$-liveness, $\betaF$-finality, and $\betaA$-accountable safety are defined analogously to their definitions in
Section~\ref{sec:model}.
Using this model, we restate the blockchain CAP theorem (\cite[Theorem 4.1]{blockchain-cap-theorem}):
\begin{proposition}[From~\cite{blockchain-cap-theorem}, see also~\cite{captheorem}]
    \label{prop:cap-theorem}
    No protocol provides both $\betaF$-finality and $\betaL$-liveness for any $\betaF, \betaL \geq 0$ under $\AdvEnvPG$.
\end{proposition}
One readily obtains the availability--accountability dilemma of~\cite{aadilemma} as a corollary of Theorem~\ref{thm:acc-implies-fin} and Proposition~\ref{prop:cap-theorem}.
\begin{corollary}
    \label{cor:aa-dilemma}
    No protocol provides both $\betaA$-accountable safety and $\betaL$-liveness for any $\betaA > 0,\betaL \geq 0$ under $\AdvEnvPG$.
\end{corollary}

\section{Finality, Accountable Safety, and Security under Partial Synchrony}
\label{sec:finality-psync}

We clarify 
the relations among 
finality, accountable safety, and security under partial synchrony.
Specifically, even though $(f+1)$-accountable safety implies $f$-finality, 
it does not imply security under partial synchrony.
To illustrate this, we consider $n, f$ with $n=3f+1$, and construct a protocol called SyncFin that is $(f+1)$-accountable safe (and $f$-safe under partial synchrony by Theorem~\ref{thm:acc-implies-fin}), yet cannot recover liveness after $\GST$ under partial synchrony, even though it is $f$-live under synchrony (the largest possible liveness resilience, \cf~Corollary~\ref{cor:acc-implies-fin-1}).

The SyncFin protocol consists of
(i)~an underlay consensus protocol executed by the $n$ replicas and secure under synchrony (\eg, Sync HotStuff~\cite{synchotstuff}, Sync-Streamlet~\cite{streamlet}), and 
(ii)~an add-on `gadget' of `finality signatures'
on the ledgers output by the underlay protocol.
The gadget works as follows:
Once for the first time at some height $h$ a block is confirmed by the underlay protocol in the view of a replica, the replica `votes for' that respective chain in the gadget
by broadcasting 
a \emph{finality signature} on the block to all other replicas.
A replica creates at most one finality signature per height, on the first block observed to be confirmed at that height by the underlay protocol.
If it later observes a conflicting block become confirmed by the underlay protocol at the same height, it does not sign that block (or any descendent thereof).
Clients confirm a block of this new protocol that is a composite of the underlay and finality-signature gadget, 
upon observing $2f+1$ finality signatures on a block and 
its prefix.
A similar add-on was contemporaneously used
for unrelated problems in~\cite{oflex,babylonlightpaper}.
\begin{theorem}
\label{thm:security}
SyncFin is $(f+1)$-accountable safe
and $f$-live under synchrony.
\end{theorem}
\begin{proof}[Sketch]
If clients output conflicting ledgers, they must have observed conflicting blocks at the same height, each with $2f+1$ finality signatures.
Since signing different blocks at the same height is a protocol violation, the forensic algorithm then identifies $f+1$ adversarial replicas by inspecting the double-signers.
Thus, SyncFin is $(f+1)$-accountable safe.
As the underlay protocol is $f$-safe and $f$-live under synchrony, SyncFin is live under synchrony if there are $2f+1$ or more honest replicas sending finality signatures.
\qed
\end{proof}

The SyncFin protocol is, however, not live under partial synchrony with any resilience.
Before $\GST$, the adversary can,
with the help of a single adversary replica, cause two honest replicas to confirm conflicting, different blocks in the underlay, and send finality signatures on their respective blocks.
After signing the blocks, the honest replicas refuse to sign any conflicting block, implying that even after $\GST$, no block is guaranteed to receive finality signatures from $2f+1$ replicas.
Hence, SyncFin is not live after $\GST$.

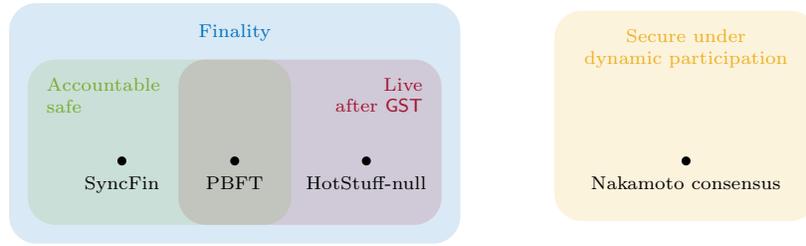
\begin{figure}[tb]
    \centering
    \begin{tikzpicture}[x=1cm,y=1cm]
        \scriptsize

        \draw [fill=myParula01Blue,fill opacity=0.15,rounded corners=1.3em,draw=none] ([yshift=0.9cm]-3cm,-2cm) rectangle ([yshift=0.1cm]3cm,2cm);
        \draw [fill=myParula07Red,fill opacity=0.15, rounded corners=1.3em, draw=none] ([yshift=0.6cm]-0.75cm,-1.45cm) rectangle ([yshift=0.4cm]2.75cm,0.95cm);
        \draw [fill=myParula05Green,fill opacity=0.15, rounded corners=1.3em, draw=none] ([yshift=0.6cm]-2.75cm,-1.45cm) rectangle ([yshift=0.4cm]0.75cm,0.95cm);

        \draw [fill=myParula03Yellow,fill opacity=0.15, rounded corners=1.3em, draw=none] ([yshift=0.4cm]4.25cm,-1.2cm) rectangle ([yshift=0.8cm]7.75cm,1.2cm);

        \node [align=center,anchor=south,yshift=1.5cm,myParula01Blue] at (0,0) {Finality};
        \node [align=right,anchor=north east,yshift=1.25cm,myParula07Red] at ([yshift=0.2cm,xshift=1.6cm]1,-0.25) {Live\\after $\GST$};
        \node [align=left,anchor=north west,yshift=1.25cm,myParula05Green] at ([yshift=0.2cm,xshift=-1.6cm]-1,-0.25) {Accountable\\safe};
        \node [align=center,anchor=south,yshift=1.15cm,myParula03Yellow] at (6,0) {Secure under\\dynamic participation};

        \node [circle,draw=none,fill=black,inner sep=1.2pt] at (0,0) {};
        \node [anchor=north,align=center,yshift=-3pt] at (0,0) {PBFT};

        \node [circle,draw=none,fill=black,inner sep=1.2pt] at (-1.5,0) {};
        \node [anchor=north,align=center,yshift=-3pt] at (-1.5,0) {SyncFin};

        \node [circle,draw=none,fill=black,inner sep=1.2pt] at (+1.75,0) {};
        \node [anchor=north,align=center,yshift=-3pt] at (+1.75,0) {HotStuff-null};

        \node [circle,draw=none,fill=black,inner sep=1.2pt] at (+6,0) {};
        \node [anchor=north,align=center,yshift=-3pt] at (+6,0) {Nakamoto consensus};
        
    \end{tikzpicture}
    \caption{Venn diagram of protocols satisfying finality, accountable safety, security under partial synchrony, and dynamic participation. The key Theorem~\ref{thm:acc-implies-fin} of this work means that accountable safe protocols are contained in the set of final protocols.}
    \label{fig:sets}
\end{figure}

We summarize the relation among protocols that satisfy finality, accountable safety, security under partial synchrony, and security under dynamic participation in Figure~\ref{fig:sets}.
The blue set contains protocols with $n = 3f+1$ replicas that are $f$-final.
The green set contains $f+1$-accountable safe protocols,
whereas the red one contains protocols that,
in addition to being $f$-final,
are also
$f$-live \emph{after $\GST$},
\ie, they are $f$-secure under partial synchrony.
Since $(f+1)$-accountable safety implies $f$-finality (Theorem~\ref{thm:acc-implies-fin}), the green set is within the blue one.
By Proposition~\ref{prop:cap-theorem}, no protocol is $\betaF$-final and $\betaL$-live under a dynamically available network for any $\betaF,\betaL \geq 0$, \ie, the blue and the yellow sets do not intersect
(and as a consequence, the green and yellow sets do not intersect---the availability--accountability dilemma, Corollary~\ref{cor:aa-dilemma}).
Finally, Theorem~\ref{thm:security} shows that SyncFin is $f$-accountable safe and $f$-live under synchrony, but as we have seen above, it is not $f$-live after $\GST$ under partial synchrony, so it is not in the red set.
PBFT~\cite{pbft} is both $f$-accountable safe~\cite{bftforensics} and $f$-safe and live under partial synchrony.
An example of a protocol that is not accountable safe, yet secure under partial synchrony is HotStuff-null~\cite{bftforensics}.

\section*{Acknowledgment}
We thank
Orfeas Stefanos Thyfronitis Litos,
David Mazières,
and
Srivatsan Sridhar
for 
fruitful discussions.
JN, ENT and DT are supported by 
Ethereum Foundation and Input Output Global.
JN is supported by the Protocol Labs PhD Fellowship.
ENT is supported by the Stanford Center for Blockchain Research.%
\bibliographystyle{splncs04}
\bibliography{references}

\end{document}